\documentclass[sigconf, dvipsnames]{acmart}

\usepackage{multirow}
\usepackage{array}
\usepackage{subfigure}
\usepackage{mdframed}
\usepackage{float}
\usepackage{enumitem}
\usepackage{breakurl} 
\usepackage{xspace}
\usepackage{amsmath}

\newcommand{\new}[1]{{#1}}
\newcommand{\final}[1]{\textcolor{black}{#1}}

\newcommand{\redacted}[1]{#1}
\newcommand*{\studyone}{\textbf{\textsc{STUDY 1}}\xspace}
\newcommand*{\studytwo}{\textbf{\textsc{STUDY 2}}\xspace}
\newcommand{\pvalue}[2]{$#1$$\times$$10^{-#2}$}
\newcommand{\bpvalue}[2]{$\mathbf{#1}$$\mathbf{\times}$$\mathbf{10^{-#2}}$}
\newcommand{\showURL}[1]{\unskip}
\newcommand{\showDOI}[1]{\unskip}

\AtBeginDocument{%
  \providecommand\BibTeX{{%
    \normalfont B\kern-0.5em{\scshape i\kern-0.25em b}\kern-0.8em\TeX}}}

\copyrightyear{2021}
\acmYear{2021}
\setcopyright{iw3c2w3}
\acmConference[WWW '21]{Proceedings of the Web Conference 2021}{April 19--23, 2021}{Ljubljana, Slovenia}
\acmBooktitle{Proceedings of the Web Conference 2021 (WWW '21), April 19--23, 2021, Ljubljana, Slovenia}
\acmPrice{}
\acmDOI{10.1145/3442381.3450093}
\acmISBN{978-1-4503-8312-7/21/04}

\def\plaintitle{Online Mobile App Usage as an Indicator of Sleep Behavior and Job Performance}
\def\plainauthors{Chunjong Park, Morelle Arian, Xin Liu, Leon Sasson, Jeffrey Kahn, Shwetak Patel, Alex Mariakakis, and Tim Althoff}
\def\plainkeywords{mobile app interaction, interaction time, sleep tracking, sleep behavior, job performance}

\begin{document}

\title{\plaintitle}


\author{Chunjong Park$^{*}$, Morelle Arian$^{*}$, Xin Liu, Leon Sasson$^{\dagger}$, Jeffrey Kahn$^{\dagger}$ \\Shwetak Patel, Alex Mariakakis${^{\scriptscriptstyle \ddagger}}$, Tim Althoff}
\thanks{{}$^{*}$Both authors contributed equally to this research.}
\affiliation{%
  \institution{University of Washington, Rise Science Inc.$^{\dagger}$, University of Toronto$^{\ddagger}$}}

%
\renewcommand{\shortauthors}{Park and Arian, et al.}
\renewcommand{\authors}{\plainauthors}

\begin{abstract}

Sleep is critical to human function, mediating factors like memory, mood, energy, and alertness; therefore, it is commonly conjectured that a good night's sleep is important for job performance. 
However, both real-world sleep behavior and job performance are difficult to measure at scale.
In this work, we demonstrate that people's everyday interactions with online mobile apps can reveal insights into their job performance in real-world contexts.
We present an observational study in which we objectively tracked the sleep behavior and job performance of salespeople ($N=15$) and athletes ($N=19$) for 18 months, leveraging a mattress sensor and online mobile app to conduct the largest study of this kind to date.
We first demonstrate that cumulative sleep measures are significantly correlated with job performance metrics, \new{showing that an hour of daily sleep loss for a week was associated with a 9.0\% average reduction in contracts established for salespeople and a 9.5\% average reduction in game grade for the athletes}. 
We then investigate the utility of online app interaction time as a passively collectible and scalable performance indicator.
We show that app interaction time is correlated with the job performance of the athletes, but not the salespeople.
To support that our app-based performance indicator truly captures meaningful variation in psychomotor function as it relates to sleep and is robust against potential confounds, we conducted a second study to evaluate the relationship between sleep behavior and app interaction time in a cohort of 274 participants.
Using a generalized additive model to control for per-participant random effects, we \new{demonstrate that participants who lost one hour of daily sleep for a week exhibited average app interaction times that were \final{5.0\%} slower.}
We also find that app interaction time exhibits meaningful chronobiologically consistent correlations with sleep history, time awake, and circadian rhythms.
The findings from this work reveal an opportunity for online app developers to generate new insights regarding cognition and productivity.
\end{abstract} 


\settopmatter{printacmref=false} 

\maketitle
 
\section{Introduction}
\label{sec:intro}
Sleep is essential to human function, affecting memory~\cite{Walker2005}, energy~\cite{Chervin2000}, mood~\cite{Brendel1990}, and alertness~\cite{Akerstedt1997}.
The importance of sleep is widely accepted, yet a significant portion of the population does not get sufficient sleep at night~\cite{Knutson2010}, and an increasing number of people report experiencing sleep problems~\cite{Basner2007}. 
In recent years, sleep tracking has become more commonplace with the introduction of commercially available sleep-tracking technologies like smartphones, smartwatches, mattress sensors, and other devices~\cite{Ko2015}. 
Online mobile apps associated with such devices collect and manage sleep data so that users can learn about and improve upon their sleep behavior. 
In doing so, many people hope to feel more rested and be more productive at their workplace.

The impact of sleep on people's psychomotor function has been widely studied, usually in controlled lab settings.
For example, researchers have found that partial sleep deprivation over multiple days can affect people's ability to perform simple tasks like reaction time tasks such as the psychomotor vigilance test (PVT) and mental math~\cite{Lo2016,VanDongen2003}.
The consequences of sleep deprivation have even been found to be comparable to the cognitive and motor impairments experienced during alcohol intoxication \cite{Williamson2000}.
Although prior literature suggests that poor sleep behavior can impact real-world job performance, this relationship has remained largely unquantified due to the lack of objective measures of both sleep behavior and job performance.
Many careers involve a complex combination of cognitive and psychomotor tasks, so it is unclear how contrived tasks like the PVT translate to higher level performance.
Furthermore, job performance assessment can require privacy-invasive methods that disrupt a person's work.
 
\final{Prior literature has leveraged technology interaction patterns as passive and scalable indicators of alertness and other aspects of psychomotor function~\cite{Murnane2016,Abdullah2014,Gordon2019,Mark2014,Pielot2015,Oulasvirta2012}.
In our work, we extend this literature by investigating the relationships between smartphone app-based performance, objective sleep behavior metrics, and objective job performance metrics gathered from two concurrent studies carried out over 18 months.}
In our first study (\studyone), we recruited 34 employees from two organizations---a bankruptcy law firm consultancy ($N=15$) and the National Football League ($N=19$)---to track their sleep using a mattress sensor, and we compared that data against widely accepted job performance metrics that were provided by their employers.
This represents is the largest study to date among those that leverage objective measurements of natural sleep behavior and job performance patterns, with similar studies including less than half the study population and only a single job category~\cite{Mah2011,Watson2017}.
Across 300 nights of sleep with associated job performance metrics, our analyses support the hypothesis that cumulative sleep loss (e.g., sleep debt, sleep history) is correlated with decreased job performance.
We find that one hour of reduced time-in-bed daily for one week was associated with 9.0\% fewer contracts established for the average salesperson and a 9.5\% grade drop for the average athlete (Section~\ref{sec:rq1}).

Since job performance metrics can be difficult to capture in practice, we explore the possibility of using timed interactions with the sleep-tracking app as an unobtrusive indicator of broader psychomotor performance.
We examined the amount of time participants spent interpreting the information on the app's main screen as an instantiation of an app-based performance metric.
We find that app interaction times were correlated with the athletes' game performance ($\rho$=-0.296, $p$=0.046), but not the salespeople's performance ($\rho$=-0.0752, $p$=0.4106) (Section~\ref{sec:rq2}). 

Although \studyone is larger than its predecessors, job performance can be extremely diverse and subject to many confounds that are infeasible to track (e.g., external personal issues leading to diminished performance).
To corroborate the use of app-based performance as a valid indicator of psychomotor function via its relationship to sleep, we tracked the sleep behavior and app interaction times of 274 individuals (\studytwo).
We analyzed this data to determine whether app interaction time is sensitive to known sleep-related influences on psychomotor function while being robust to other individual-level effects like user-specific baselines and smartphone specifications.
Across 7,200 tracked nights of sleep with more than 16,000 app interaction events, our analyses reveal that daily variations in app interaction time aligned with constructs in sleep biology, most notably circadian rhythm~\cite{Dijk1992,Borbely2016,Akerstedt1997} and sleep inertia~\cite{Akerstedt1997}.
We find that app interaction time was negatively correlated with time-in-bed ($\rho$=-0.015, $p$=0.049), sleep history ($\rho$=-0.055, $p$=\pvalue{3.9}{11}), and sleep debt ($\rho$=-0.095, $p$=\pvalue{5.2}{30}).
Furthermore, we demonstrate that participants who lost one hour of daily sleep over the previous week exhibited app interaction times that were 0.5 seconds slower (Section~\ref{sec:rq3}). 
In summary, our research investigates the following questions:
\begin{enumerate}[leftmargin=0.03\linewidth, topsep=0pt,itemsep=0pt,parsep=0pt,partopsep=0pt]
        \item[] \textbf{RQ.1} Is sleep behavior correlated with job performance?\\ (\studyone, Section~\ref{sec:rq1})
        \item[] \textbf{RQ.2} Is app-based performance correlated with job performance? (\studyone, Section~\ref{sec:rq2}) 
        \item[] \textbf{RQ.3} Is app-based performance correlated with sleep behavior? (\studytwo, Section~\ref{sec:rq3})
\end{enumerate}
\section{Related Work}
\label{sec:related}
In this section, we describe prior work on (1) sleep biology, (2) consumer sleep-tracking applications and their effects on users' sleep behavior, (3) the relationship between sleep and performance, and (4) the use of technology to passively infer performance. 

\subsection{Models in Sleep Biology}
\label{sec:related-sleepbio}
Dijk et al.~\cite{Dijk1992,Borbely2016} describe sleep biology using an additive two-process model consisting of \textit{circadian rhythm}, the 24-hour biological cycle that occurs in nearly all creatures, and \textit{homeostasis}, the increasing pressure to sleep as one stays awake for longer periods of time.
Akerstedt et al.~\cite{Akerstedt1997} later added a third process: \textit{sleep inertia}, the initial drowsiness that occurs immediately after waking up.
Past studies have used two- and three-process models of sleep biology to understand the effects of sleep schedules on mood~\cite{Golder2011} and athletic performance~\cite{Thun2015}.
Genetic predispositions and chronotyping (i.e., ``morning person'' vs. a ``night owl'') have also been shown to affect sleep behavior~\cite{Roenneberg2003,Allebrandt2010}.
Matchock et al.~\cite{Matchock2009} and Althoff et al.~\cite{Althoff2017} find evidence of significant interaction between circadian rhythms and chronotyping on reaction times.
A recent study suggests that female sleep duration is not strongly dependent on menstrual cycles~\cite{pierson2021daily}.

To characterize the importance of sleep, many studies require that subjects adhere to strict sleep schedules that range from a full night's rest to total sleep deprivation \cite{Pilcher1996}; however, researchers have noted that natural sleep is more commonly characterized by partial sleep deprivation over multiple days, also known as chronic sleep restriction.
Regardless of the specifics of one's sleep schedule, researchers have noted the accumulation of sleep debt as an important metric of sleep behavior~\cite{VanDongen2003,Dinges2004,Horne2004,Dinges1997}.
Calculating sleep debt requires understanding an individual's sleep need---the optimal amount of daily sleep an individual requires.
Sleep need is often measured through a controlled study where a participant is subjected to extended time-in-bed over many days;
under these conditions, sleep length typically decays exponentially over time and approaches an asymptote that represents the individual's sleep need~\cite{Klerman2008}. 
Since sleep need is often difficult to measure in uncontrolled settings, Kitamura et al.~\cite{Kitamura2016} propose a method of estimating sleep debt based on one's history of time-in-bed.
We utilize Kitamura et al.'s calculation of sleep debt as a cumulative sleep metric in our analyses.
Furthermore, we leverage Akerstedt et al.'s three-process model to understand the correlations between sleep debt, job performance, and our app-based performance metric.

\subsection{Sleep Sensing Systems}
\label{sec:related-sleepsystems}
Traditional polysomnography studies utilize expensive sensors like EEGs and EMGs to get fine-grained information about how a person sleeps~\cite{Ibanez2018}.
Given the growing desire for health-related self-tracking technologies, sleep tracking has become more commonplace.
Ko et al.~\cite{Ko2015} provide a review of consumer sleep sensing technologies.
Their review covers sleep-sensing form factors like smartphones \cite{Min2014,Chen2013}, smartwatches, wristbands, mattress sensors, and wireless radios \cite{Rahman2015}. 
Overall, these technologies are able to gather sleep metrics ranging from sleep duration to disturbance frequency.
For instance, Min et al. \cite{Min2014} propose a mobile app that processes seven different sensor streams (e.g., motion, sound, light) to classify a person's sleep state and sleep quality, while
Rahman et al. \cite{Rahman2015} demonstrate that coarse body movements and subtle chest movements from breathing and heartbeats can be detected by measuring the reflections of high-frequency wireless signals.

Beyond exploring new ways of extracting sleep data, another line of research has explored how sleep data should be presented to users and how recommendations should be generated to improve sleep behavior. 
Bauer et al.~\cite{Bauer2012} show that a recommendation-based peripheral display can serve as a low-effort, yet effective, method for improving awareness of healthy sleep behavior.
Daskalova et al.~\cite{Daskalova2016} address the creation of personalized recommendations through guided self-experimentation.
Their mobile app, SleepCoacher, tracks sleep behavior metrics from the accelerometer and microphone to generate data-driven recommendations; as users engage with these recommendations, SleepCoacher is able to measure whether the intervention had its intended effect. 
Daskalova et al. have also explored cohort-based sleep tracking and recommendations~\cite{Daskalova2018}.

To the best of our knowledge, the aformentioned body of literature has not explored the opportunity of using interactions with a sleep-tracking system as an additional source of information.
The act of examining sleep data on a smartphone requires users to exert cognitive load, which itself is tied to sleep behavior.
We introduce the notion of app-based performance to investigate whether interactions with a sleep-tracking system can provide insight into sleep behavior or job performance.
\vspace{-3.00mm}

\subsection{The Effects of Sleep on Performance}
\label{sec:related-sleepperformance}
In this work, we contextualize large-scale sleep data through job-based and mobile app-based performance measurements~\cite{althoff2017population}.
One of the most common tests that have been used for measuring psychomotor function is the psychomotor vigilance test (PVT)~\cite{Dinges1985,Jewett1999,Belenky2003}, during which a person is asked to respond to a visual signal by pressing a button.
Researchers have employed a variety of other contrived tasks to measure cognitive and motor performance in relation to sleep.
Pilcher and Huffcutt~\cite{Pilcher1996} provide a meta-analysis of 19 research studies that examine task performance and mood as a function of sleep restriction.
The cognitive tasks Pilcher and Huffcut list in their review include logical reasoning~\cite{Blagrove1995}, mental math~\cite{Baranski1997}, visual search tasks~\cite{Baranski1997}, and word memory tasks~\cite{Mazzoni1999}.
The motor tasks they reference include exercise~\cite{Pickett1975,MartinGaddis1981}, endurance tasks~\cite{Martin1981}, and muscle strength tests~\cite{VanHelder1989}.

The aforementioned tests have been used to examine the effects of sleep quality on various performance dimensions.
Rajdev et al.~\cite{Rajdev2013} use the PVT to validate a mathematical model of psychomotor performance based on sleep debt.
Ramakrishnan et al.~\cite{Ramakrishnan2012} also use the PVT to validate their own phenotype-specific group-average model of psychomotor performance.
Lo et al.~\cite{Lo2016} use a battery of seven cognitive tasks to find that partial sleep deprivation impairs a wide range of cognitive functions, subjective alertness, and mood.
Lastly, Killgore et al.~\cite{Killgore2006,Killgore2008} study the effects of total sleep deprivation on measures of emotional intelligence, constructive thinking, and decision-making during a gambling task.

Watson~\cite{Watson2017} provides a literature review on the interaction between sleep and athletic performance, drawing closer to our research questions on sleep and job performance.
In one of the most closely related studies to our own \studyone, Mah et al.~\cite{Mah2011} examined the effects of sleep on collegiate basketball players.
Their study entailed having athletes maintain their typical sleep schedule, after which they underwent a period of sleep extension with a minimum goal of 10 hours in bed each night.
The athletes were evaluated using the PVT, subjective scales of sleepiness, and performance metrics specific to basketball practices (e.g., sprint time, shooting percentage).
Furthermore, the athletes were asked to rate their own performance during practices and games.
Although research like that of Mah et al. strives towards our goal of measuring high-level job performance rather than low-level task performance, their work falls short capturing an objective measurement of in-game performance.
In our work, we build on this literature by examining the correlation between sleep behavior and natural, widely accepted job performance metrics collected from the workplaces of athletes and salespeople (\studyone).
Our study represents the largest effort to study this relationship to date, spanning multiple careers categories and nearly 300 nights of data in a study population that more than doubles prior work~\cite{Mah2011}.

\subsection{Interactions as an Indicator of Cognition}
\label{sec:related-interaction}
Technology interaction patterns have been used as an indicator for understanding different aspects of performance.
Regarding smartphones, indicators like app usage and app-specific productivity have been used to estimate alertness~\cite{Murnane2016,Abdullah2014,Gordon2019,Mark2014,Pielot2015,Oulasvirta2012}.
For example, Murnane et al.~\cite{Murnane2016} demonstrate that app usage patterns vary for individuals with different chronotypes and rhythms of alertness.
Oulasvirta et al.~\cite{Oulasvirta2012} correlate frequent, short bursts of smartphone interaction (i.e., checking a notification) with inattentiveness.
Other researchers have leveraged technology interaction to estimate higher level constructs like stress~\cite{Ferdous2015}, mood~\cite{Golder2011,Mehrotra2016}, academic performance~\cite{Wang2015}, inebriation~\cite{Bae2017,Mariakakis2018}, and accident risk~\cite{Althoff2018}.

Even with just the timing between two interaction events, researchers have been able to assess aspects of a person's cognition.
Vizer et al.~\cite{Vizer2009} analyze variation in computer keystroke rate to infer increased stress levels.
Althoff et al.~\cite{Althoff2017} track users' typing and click speed on a web search engine as a measure of psychomotor function.
Their work shows that keystroke time and click time vary based on sleep duration, circadian rhythms, and the homeostatic process.
Inspired by this prior work, we measure app interaction time as a potential non-intrusive indicator of performance (\studyone) and sleep (\studytwo). 
Concretely, we demonstrate that this indicator correlates with athletic job performance and meaningfully reflects psychomotor performance variation due to biological functions (i.e., circadian rhythm and sleep inertia).
\section{Performance and Sleep Data}
\label{sec:dataset}
Both of our observational studies, \studyone and \studytwo, followed the same protocol.
In this section, we first describe the technology that participants used to track and monitor their sleep behavior.
We then describe the metrics that we use to quantify sleep behavior, job performance, and app-based performance.
We conclude by detailing the procedures that were used to clean the dataset in preparation for analysis.

\begin{figure}
    \centering
    \includegraphics[height=6cm]{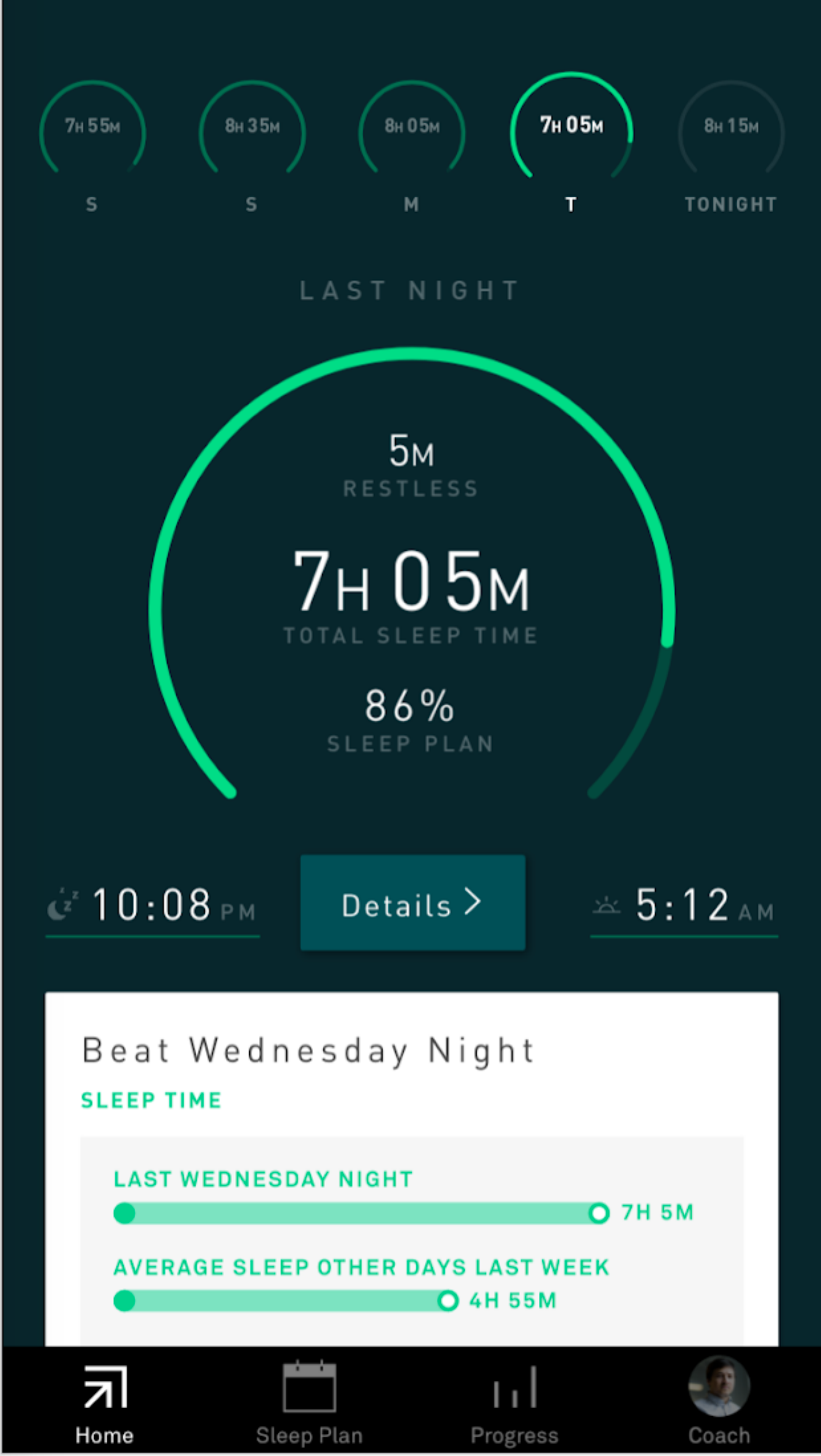}
    \caption{The home screen of \redacted{Rise Science's} sleep-tracking app shows data about the user's most recent night of sleep.}
    \Description[The home screen of a sleep-tracking app.]{The shows the home screen of a sleep-tracking app. The top of the screen shows summary statistics for the user's previous night of sleep, while insights from the user's history are shown below.}
    \label{fig:app-home-screen}
    \vspace{-5.00mm}
\end{figure}

\subsection{Devices}
\label{sec:dataset-devices}
Participants in both \studyone and \studytwo were recruited as existing users of \redacted{Rise Science's}\footnote{\redacted{\url{https://www.risescience.com/}}} mobile app.
We enrolled participants via targeted recruitment (salespeople and athletes) as well as broader recruitment calls, but participants from all sources were onboarded in a similar fashion.
Each participant \new{received} a kit consisting of an Emfit QS\footnote{\url{https://qs.emfit.com/}} and a sleep-tracking mobile app\footnote{\redacted{\url{https://play.google.com/store/apps/details?id=com.risesci.risesciapp}}}\textsuperscript{,}\footnote{\redacted{\url{https://apps.apple.com/us/app/rise-science/id1107659850?app=itunes&ign-mpt=uo\%3D4}}} as shown in Figure~\ref{fig:app-home-screen}.
The Emfit QS is a highly sensitive pressure sensor that lies underneath the user's mattress (or their preferred side of the mattress when the bed is shared).
The sensor uses ballistocardiography to track heart rate, breathing rate, and movement.
In past studies, the Emfit QS has been validated against a standard clinical heart rate monitor and polysomnography equipment~\cite{Ranta2019,Guerrero-Mora2012,Kortelainen2010}.
Within the sleep-tracking mobile app, participants can access and visualize their own sleep data, view sleep session summaries, create sleep plans, and learn about the importance of sleep.

\subsection{Participants and Study Procedure}
The data collection period started on May 2017 and ended on December 2018, spanning 592 days.
Recruitment happened throughout that period, and participants joined and left the study at their own discretion.
In \studyone (Section~\ref{sec:study1}), participants from the bankruptcy law firm consultancy were enrolled for 225 days, and participants from the NFL teams were enrolled for 450--520 days. 
In \studytwo (Section~\ref{sec:study2}), participants were enrolled for 80--580 days.
Demographic data like age and gender were not collected to minimize intrusion and maintain privacy. 
However, we know that most NFL players are between 23--27 years old~\cite{Gertz2017}, and a public database estimated that 71\% of the bankruptcy law firm's employees are millenials\footnote{The firm's name is removed to preserve their anonymity.}. 
In \studytwo, we expect that most of the participants were under the age of 45 since self-tracking requires active engagement with sleep-technology, which is more common in younger demographics~\cite{cea2015sleep}.

\final{
Special care was taken to avoid coercion during recruitment. 
To protect participants' privacy, employers were never told who enrolled in the study and were only given aggregated results after the study was done.
}
Participants did not receive explicit instructions from the research team and were free to follow whatever sleep schedule they chose. 
Participants were also free to use the Emfit QS and sleep-tracking app at will; if they had to travel while participating in the study, they could choose to either bring the Emfit QS with them or leave it behind.
The mobile app sent participants notifications, reminders, and recommendations for improving their sleep (e.g., reducing caffeine intake, dimming lights); participants were free to disable these features at any time. 
Our retrospective data analysis was conducted in accordance with the Institutional Review Board at \redacted{the University of Washington}.

\begin{table*}[t]
\centering
\begin{tabular}{|p{2cm}|p{3.5cm}|p{9.6cm}|}
\hline
& \textbf{Metric} & \textbf{Description} \\ \hline
\multirow{6}{*}{\textbf{Sleep}} & Bedtime & Time at which the user got into their bed \\ \cline{2-3} 
 & Wake time & Time at which the user got out of their bed \\ \cline{2-3} 
 & Midpoint & Midpoint between start and end time \\ \cline{2-3} 
 & Time-in-bed & The total time the user spent in bed during a single day including nighttime sleep and naps, regardless of whether they were sleeping \\ \cline{2-3} 
 & Sleep debt & Weighted average of difference between sleep need and time-in-bed \\ \cline{2-3} 
 & Sleep history & Weighted average of time-in-bed \\ \hline
\multirow{2}{*}{\textbf{\begin{tabular}[c]{@{}l@{}}Job\\ Performance\end{tabular}}} & Number of hires (salespeople) & Number of contracts made after consulting, normalized by the number of hours they work \\ \cline{2-3} 
 & Game grade (athletes) & Score of a player's game performance out of 100 assigned through three independent experts \\ \hline 
\textbf{App Usage} & Interaction time & Time between opening home screen of app to another screen by user's touch input \\ \cline{2-3} 
\hline
\end{tabular}
\caption{A summary of the metrics we collect in our dataset through three data streams: (1) sleep metrics through the Emfit QS, (2) job performance through the participants' employers, and (3) app usage through a sleep-tracking mobile app.}
\label{tab:metrics}
\vspace{-7.00mm}
\end{table*}

\subsection{Sleep Behavior Metrics}
\label{sec:dataset-sleep}
The Emfit QS reports the following metrics to describe a single night's rest: bedtime, wake time, sleep midpoint, time-in-bed, and total sleep duration.
Time-in-bed measures how long a person is in their bed, thus only requiring accurate presence detection.
Total sleep duration, on the other hand, estimates how long a person is actually asleep in their bed, thus requiring both accurate presence and sleep detection.
Because total sleep duration is more susceptible to sensing errors, we exclude it from the analyses reported in this paper and focus on time-in-bed measures, which can be measured with higher accuracy and validity.
This choice is common in previous work as well~\cite{Abdullah2014, Abdullah2016, walch2016global, Althoff2017,ancoli2003role}.
Nevertheless, sleep duration and time-in-bed were strongly correlated in our dataset ($\rho=0.85, p<0.001$) and produced comparable results in most cases.
Looking beyond a single night's sleep, cumulative metrics across multiple nights can provide further insight into participants' sleep behavior.
We use sleep debt~\cite{Kitamura2016,Dinges1997,VanDongen2003}, the weighted accumulation of sleep loss, as one of those measures.
Sleep debt is calculated using the following formula:
\begin{eqnarray*}
    \sum_{i=1}^{7}{-e^{-i/7} * (\text{SleepNeed} - \text{TimeInBed}_i)}
\end{eqnarray*}
where $i$ is the number of days in the past.
Note that the difference between sleep need and debt is weighted by a decaying exponential with a time constant of 7 days \cite{Ramakrishnan2016}, indicating that recent measurements have greater importance.
Whenever a participant skipped a day of tracking, we impute the missing time-in-bed value using their average time-in-bed over the past week. 
Significant imputation happened in only 12.7\% of the weeks (see Section~\ref{sec:datasets-filtering} for details).
Sleep need is typically estimated in a controlled laboratory study, making it challenging to estimate sleep debt in the wild.
Therefore, we estimate sleep need using the approach proposed by Kitamura et al.~\cite{Kitamura2016}.
Their approach involves using long nights of sleep to predict the difference between sleep need and habitual sleep (i.e., the average time-in-bed over two weeks) for a minimum of four nights. 
We also introduce a simplified \textit{sleep history} metric that avoids the notion of sleep need but still captures an aggregate measure of sleep behavior: 
\begin{eqnarray*}
    \frac{1}{\sum_{n=1}^{7}{e^{-n/7}}}
    \sum_{i=1}^{7}{e^{-i/7} * \text{TimeInBed}_i}
\end{eqnarray*}
The calculation of sleep history is normalized such that weights sum to one, making the metric more interpretable as a weighted average of time-in-bed over the past week. 

\subsection{Job Performance Metrics}
\label{sec:datasets-job}
For \studyone, we were able to utilize organizational partnerships to gather the job-specific performance metrics described below:

\subsubsection{Performance Metrics for Salespeople}
The salespeople who participated in our study work at a bankruptcy law firm consultancy.
Their job entails fielding phone calls from potential clients in need of bankruptcy relief and referring those callers to an attorney.
The employees collect a fee upon successfully hiring a client, which is the company's primary revenue source.
Employees in this company are evaluated on a variety of metrics related to that revenue stream, such as the amount they collect in fees.
However, the distribution of fees is highly variable (\$250--\$1750) and primarily dependent upon the clients rather than the employees themselves.
Therefore, we focus on the \textit{number of hires per day} the salespeople were able to establish as their job performance metric.
This metric follows a right-sided normal distribution~\cite{Leone1961} with a mean of 3.80 and a standard deviation of 3.27 hires per day.
Although work hours were generally consistent across the company, we normalized the number of hires a salesperson made by the number of hours they worked that day to account for whatever variance remained.

\subsubsection{Performance Metrics for Athletes}
The athletes who participated in our study play in a professional American football league in the United States.
We gather job performance metrics for the athletes' performance during weekly games using Pro Football Focus\footnote{\url{https://www.pff.com/}} (PFF).
PFF evaluates athletes using the following procedure~\cite{ProFootballFocus2017}: two experts score every play the athlete is involved in, a third expert resolves disagreements between those experts, an external group of ex-players and coaches verifies the scores, and then the scores are summed together and normalized to a grade between 0--100.
Although PFF is not purely quantitative, the experts can account for in-game context that is lost by purely statistical methods (e.g., injuries, matchups).
For this reason, PFF has been used in the past literature for assessing performance in football \cite{Dodson2016,Provencher2018,Byanna2016}.

In American football, each player has their own unique skill set according to their position; for example, quarterbacks are typically known for their throwing ability and wide receivers are known for their speed and catching ability.
The notion of positional specialization makes it difficult to compare athletes across positions in a purely quantitative way, especially since some skills are position-specific.
Nevertheless, PFF's method of expert observation and score normalization allows them to produce an \textit{overall game performance} grade that can be used to compare athletes across positions.

\subsection{Sleep-Tracking App Usage Metrics}
\label{sec:datasets-app}
Participants had to interact with a sleep-tracking app in order to examine their sleep summaries, so we leverage these interactions as a novel source of data.
We take inspiration from Althoff et al.~\cite{Althoff2017} by using \textit{app interaction time}---the time between two touch events in the app---as an app-based performance metric.
App interaction time is not meant to be a direct replacement of the PVT; instead, it serves as a more general measure of cognition by measuring the user's ability to process information on the app's screen.
Interaction speed can be confounded by the content that is shown on the screen.
To account for this confound, we restrict our analysis of app interaction time to transitions from the home screen (shown in Figure \ref{fig:app-home-screen}) to other endpoints within or outside of the app.

\newcolumntype{P}[1]{>{\centering\arraybackslash}p{#1}}

\begin{table*}[t]
\centering
\begin{tabular}{|p{10cm}|P{2.25cm}|P{2.25cm}|} \hline
\textbf{\studyone Statistics} & \textbf{Salespeople} & \textbf{Athletes} \\ \hline
Number of participants & 15 & 19 \\ \hline
Total unique days with both sleep-tracking and job performance measurements & 118 & 171 \\ \hline
Total unique days with both app interaction and job performance measurements & 122 & 46 \\ \hline
Total nights of sleep tracked with app-based performance measure & 234 & 418 \\ \hline
Total nights of sleep tracked & 834 & 2,687 \\ \hline
Total number of transitions between screens & 679 & 909 \\ \hline
Total number of times app was opened & 425 & 691 \\ \hline
Nights of sleep tracked per user (avg $\pm$ std) & 46.33 $\pm$ 37.45 & 133.1 $\pm$ 89.92 \\ \hline
Time-in-bed in hours (avg $\pm$ std) & 7.283 $\pm$ 2.020 & 7.308 $\pm$ 1.920 \\ \hline
Days of app use per user (avg $\pm$ std) & 28.25 $\pm$ 21.06 & 40.65 $\pm$ 50.43 \\ \hline
\end{tabular}
\caption{Summary statistics for our dataset in \studyone after \new{the filtering described in Section~\ref{sec:datasets-filtering}}.}
\label{tab:study1-dataset}
\vspace{-7.00mm}
\end{table*}

\begin{table*}[]
\begin{tabular}{cc|*3{>{\centering\arraybackslash}m{0.75in}}|*3{>{\centering\arraybackslash}m{0.75in}}|}
\cline{3-8}
 &  & \multicolumn{6}{c|}{\textbf{Sleep Metrics}} \\ \cline{3-8} 
 &  & \multicolumn{3}{c|}{\textbf{Raw Metrics}} & \multicolumn{3}{c|}{\textbf{Per-Person Z-Normalized Metrics}} \\ \cline{3-8} 
 &  & \multicolumn{1}{c}{Time-in-Bed} & \multicolumn{1}{c}{Sleep Debt} & Sleep History & \multicolumn{1}{c}{Time-in-Bed} & \multicolumn{1}{c}{Sleep Debt} & Sleep History \\ \hline
\multicolumn{1}{|c|}{\rotatebox{90}{\rlap{\hspace{-1.60cm}\tabular{@{}c}\textbf{Job} \\\textbf{Performance} \\ \textbf{Metrics}\endtabular}}}  & \multicolumn{1}{c|}{\begin{tabular}[c]{@{}c@{}}NFL Player\\ Game Grades\\ ($N$ = 19)\end{tabular}} & -0.024 ($p$=0.751) & -0.095 ($p$=0.218) & -0.029 ($p$=0.711) & 0.086 ($p$=0.263) & \textbf{0.166 ($p$=0.031)} & \textbf{0.179 ($p$=0.020)} \\ \cline{2-8} 
\multicolumn{1}{|c|}{} & \multicolumn{1}{c|}{\begin{tabular}[c]{@{}c@{}}Salespeople\\ Hires per Day\\ ($N$ = 15)\end{tabular}} & -0.067 ($p$=0.469) & \textbf{0.218 ($p$=0.022)} & 0.039 ($p$=0.690) & -0.102 ($p$=0.283) & 0.164 ($p$=0.088) & -0.047 ($p$=0.634) \\ \hline
\end{tabular}
\caption{Spearman correlation coefficients between sleep behavior and job performance. P-values are provided in parentheses; results with p-value < 0.05 are shown in bold.}
\label{tab:spearman-corr-job-perf}
\vspace{-7.00mm}
\end{table*}



\subsection{Data Filtering and Post-Processing}
\label{sec:datasets-filtering}
Sleep behavior, job performance, and app interaction metrics (Table~\ref{tab:metrics}) were collected from separate sources at different intervals.
Therefore, post-processing was needed to join and collate them.

\subsubsection{General Post-Processing}
We followed best-practices in preparing mobile app data for analysis~\cite{hicks2019best}.
The calculation of time-in-bed included naps, which were either automatically annotated if the user's bedtime or wake time fell in the afternoon (12:00-18:00) or manually annotated by the user.
Naps appeared in 9.3\% of the nightly sleep metrics (62\% automatically tagged vs. 38\% manually annotated), contributing an additional 1.22 hours to time-in-bed on average.
Sleep events when participants spent more than 16 hours in bed in a single session were attributed to faulty sensing and removed from the dataset.
The remaining nights, along with imputed averages for missing values, were used for calculating sleep debt and sleep history.
A full week of sleep data was available for calculating 46.9\% of the cumulative sleep metrics, meaning that no imputation was needed for them; three or more nights were only missing in 12.7\% of the cumulative sleep metrics.
When cumulative sleep metrics were calculated without imputation, the standard deviation of the times within the same week was only 1 hour and 10 minutes; this shows that there was not significant variance within a week, justifying the use of a short-term average.
For the analyses related to app-based performance, interaction events that were shorter than 0.45 seconds ($2.5^{\text{th}}$-percentile) were excluded since these were likely accidental or automatically generated by the app itself; events longer than 54.83 seconds ($97.5^{\text{th}}$-percentile) were excluded since they were likely indicative of the user engaging in another activity.

\subsubsection{Job-Specific Filtering}
Job performance data for the salespeople was collected on a daily basis.
Therefore, every night of sleep that a salesperson tracked with their Emfit QS was collated with the job performance metric from the next day.
Aligning the data streams for the athletes was more difficult since they had games on a weekly basis. 
The athletes also had to travel to games away from their home stadium, leaving larger gaps in their sleep-tracking data.
To accommodate these issues, we aligned the weekly PFF grades with the sleep behavior metrics from the most recent tracked night of sleep within the two nights before the relevant game day; if no nights were tracked in that span, the game grade from that week was filtered out.

\subsection{Distribution of Job Performance Data}
Using D'Agostino's $K^2$ test~\cite{DAgostino1971}, we determined that the job performance metrics in our dataset were non-normally distributed (number of hires: $K^2$=21.37, $p$=\pvalue{2.3}{5}; game grades: $K^2$=14.87, $p$=\pvalue{5.9}{4}).
The same holds true for app-based performance ($K^2=5177$, p<\pvalue{1.0}{20}) and app event count ($K^2$=71.60, $p$=\pvalue{2.8}{16}).
Therefore, we use Spearman's Rank Correlation ($\rho$) across all correlational analyses throughout this paper. 
\section{\studyone: Athletes and Salespeople}
\label{sec:study1}
\new{Our first study investigates \textbf{RQ.1} and \textbf{RQ.2} within a cohort of 15 salespeople and 19 athletes.}
Table~\ref{tab:study1-dataset} shows the summary statistics of our dataset after post-processing.
The large standard deviations in the various metrics are due to the logistics of our study.
Participants were recruited throughout the 18-month-long period, so some people had many more opportunities to use the sleep-tracking tools than others.

\subsection{RQ.1: The Relationship Between Sleep Behavior and Job Performance}
\label{sec:rq1}
\final{Using the objective job performance metrics we were able to obtain from our participants' employers, we first examine whether better sleep behavior improves job performance.
To the best of our knowledge, our study is the largest to date on this topic without any constraints on how participants slept or went about their daily jobs~\cite{Lo2016,Mah2011}.
The code for all of our analyses can be found in the GitHub repository associated with this project\footnote{\final{Code available at \url{https://github.com/cjpark87/mobile-app-sleep-performance}.}}.}

\subsubsection{Analysis Procedure}
For this analysis, we calculate correlation coefficients between the job performance metrics and three sleep behavior metrics: time-in-bed, sleep debt, and sleep history.
Sleep metrics can vary across individuals due to genetic predisposition and other possible confounds~\cite{Roenneberg2003,Allebrandt2010}, so we repeat the analysis using standardized sleep behavior metrics according the Z-score within each individual's data.
Participants who did not track at least 5 nights of sleep were excluded from this analysis to ensure that the data was representative of their typical sleep behavior.
The salespeople and athletes contributed data from 118 and 171 nights of sleep with corresponding job performance metrics, respectively.

\subsubsection{Results}
The correlation coefficients between the sleep behavior and job performance metrics in our dataset are presented in Table~\ref{tab:spearman-corr-job-perf}.
The analysis reveals positive, statistically significant correlations in some, but not all, cases.
For the salespeople, sleep debt was positively correlated with the number of hires they made ($\rho$=0.218, $p$=0.022). 
For the athletes, normalized sleep history ($\rho$=0.179, $p$=0.020) and sleep debt ($\rho$=0.166, $p$=0.031) were both positively correlated with game performance.
Fewer correlations were found for the salespeople than the athletes, which could be due to the nature of their jobs.
The athletes rely on millisecond-scale reaction times during their games, whereas salespeople do not need to operate at such a rapid pace.
These results could imply that careers focused on physical and psychomotor skills may be more strongly affected by sleep behaviors than careers that focus primarily on cognition.
The fact that multiple correlations emerged between cumulative sleep behavior metrics and job performance, combined with the lack of such correlations from single-day metrics, suggests that sleep over an extended period has a stronger impact on a person's job performance than a single night of sleep. 
Additionally, the general increase in correlation coefficients after the sleep behavior metrics were normalized within individuals supports the notion that sleep needs and behaviors vary between individuals.

\begin{figure}[t]
    \vspace{-3.00mm}
    \centering
    \subfigure[Sleep Debt]{
        \includegraphics[width=0.5\linewidth]{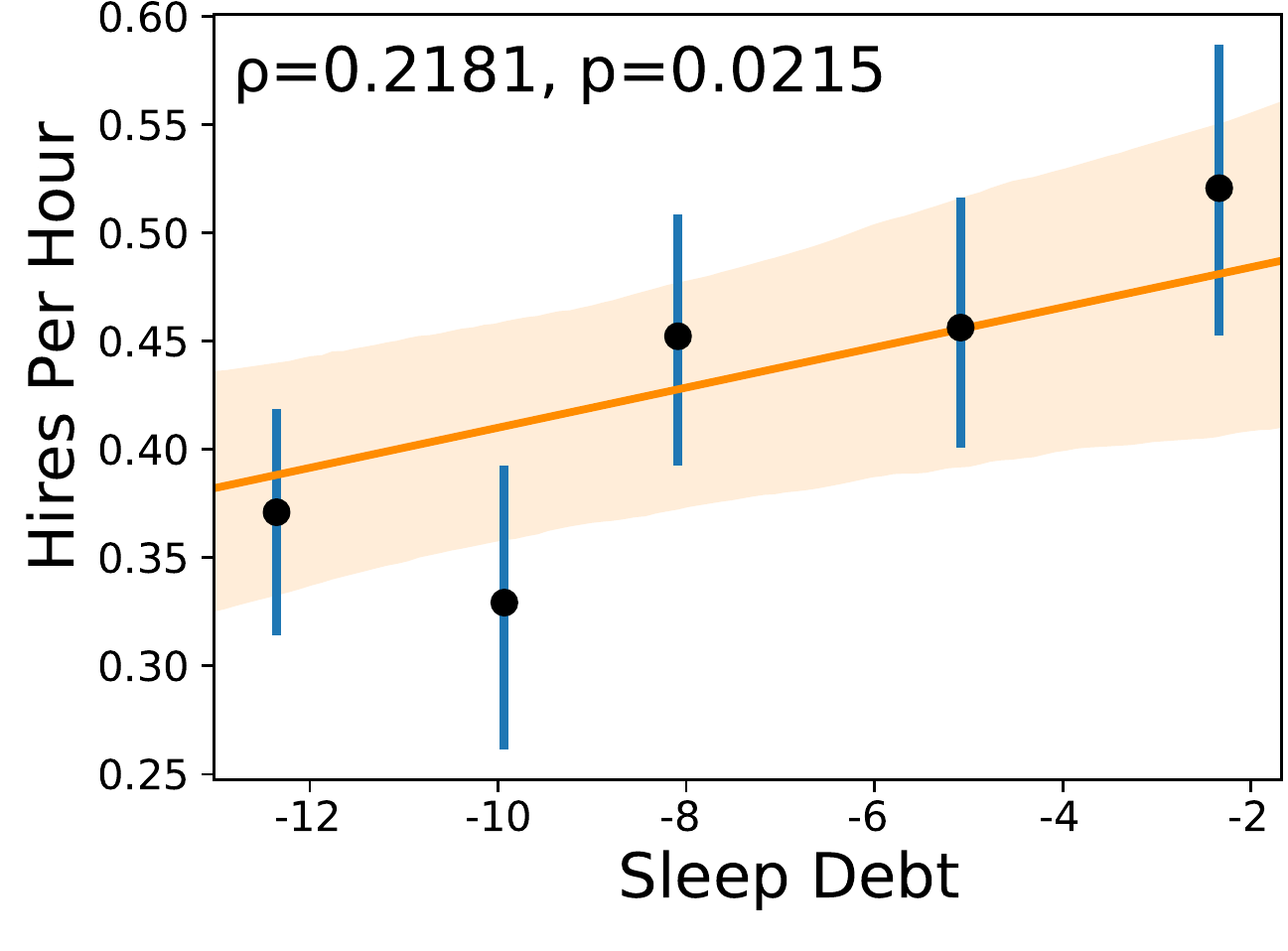}
        \label{fig:sleep-job-stat-sig-a}
    }
    \subfigure[Normalized Sleep History]{
        \includegraphics[width=0.5\linewidth]{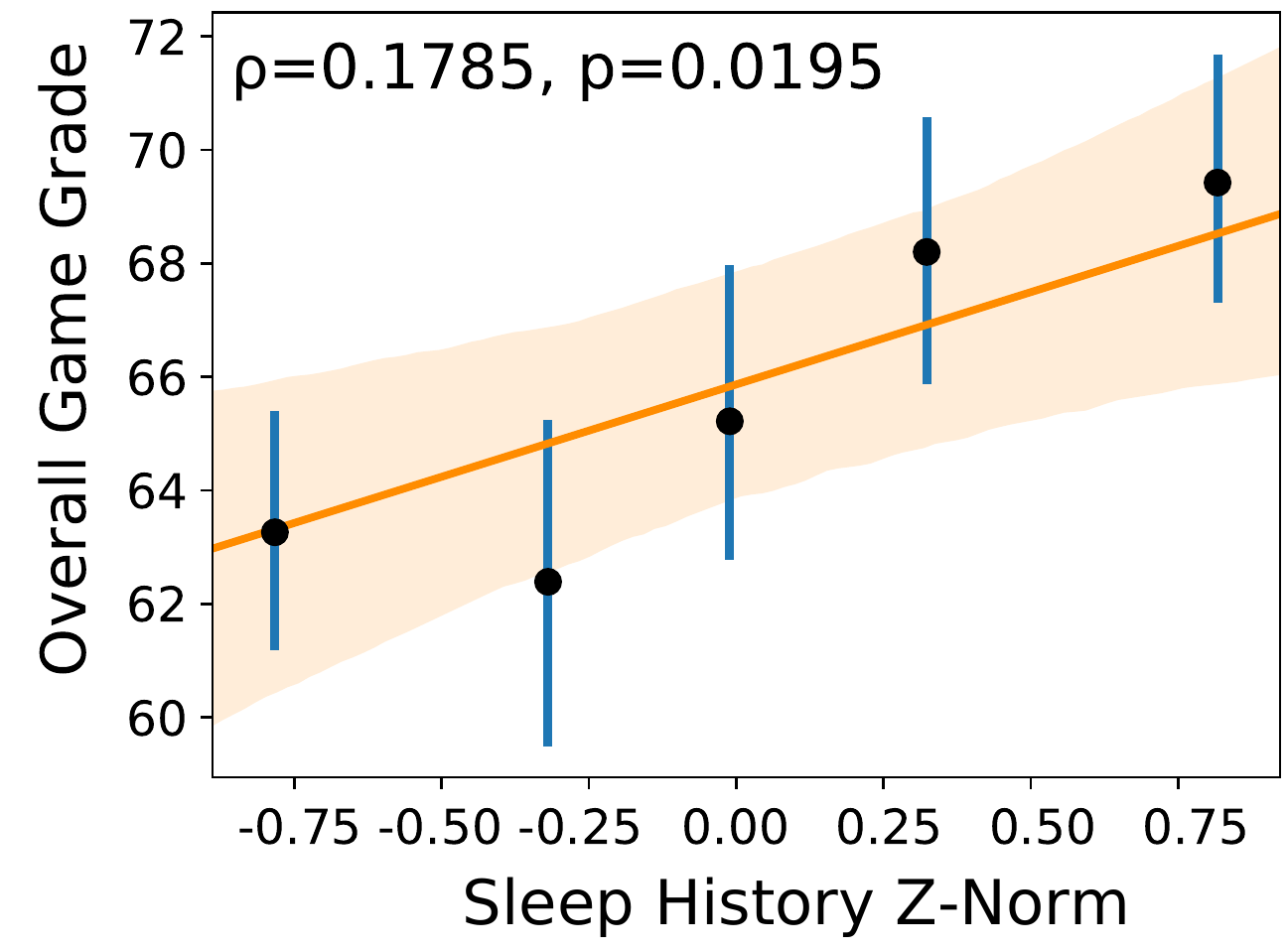}
        \label{fig:sleep-job-stat-sig-b}
    }%
    \subfigure[Normalized Sleep Debt]{
        \includegraphics[width=0.5\linewidth]{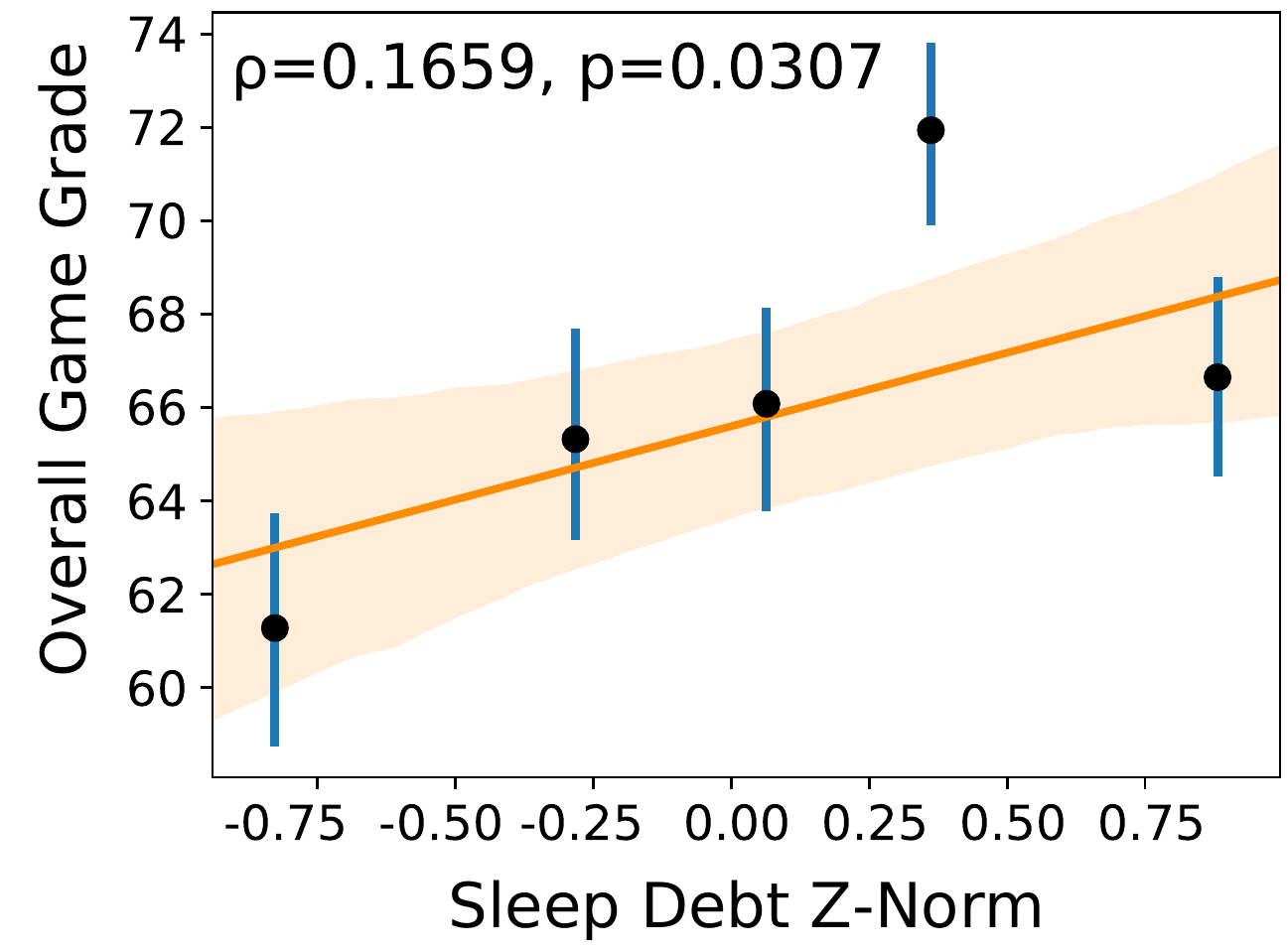}
        \label{fig:sleep-job-stat-sig-c}
    }
    \vspace{-3.00mm}
    \caption{Regression plots showing the effect sizes for the statistically significant results from Table \ref{tab:spearman-corr-job-perf}. 
    The job performance of both the (a) salespeople and (b+c) athletes is sensitive to cumulative sleep metrics.
    Throughout the paper, the data are binned into discrete and evenly distributed intervals (quintiles or deciles). 
    Point estimate and error bars represent mean estimate (black) and standard error (blue), respectively.
    Orange lines represent the best linear regression fit to the raw data along with standard errors (shaded area).
    }
    \label{fig:sleep-job-stat-sig}
    \vspace{-5.00mm}
\end{figure}
 
We further analyze the statistically significant correlations by measuring their effect sizes, which are shown in Figure~\ref{fig:sleep-job-stat-sig}. 
One hour of sleep debt by the average salesperson was associated with a 2.2\% decrease in the number of hires they were able to make.
Since sleep debt is a weighted sum of sleep deficits, another way to consider this effect size is by saying that one hour of sleep loss the night before was associated with 1.9\% fewer hires.
The average salesperson made 3.8 hires per workday and collected \$936 in fees per hire. 
Therefore, a 1.9\% decrease translates to a \$67 loss per day.
The average athlete experienced a 2.0\% drop (1.3 points) in their game grade when they lost one hour of sleep the night before.
Although these performance decreases may appear small, they can accumulate over time or across multiple people on the same team.
In fact, sleep debt implies that a deficit can be spread over multiple days, so one hour of sleep loss the night before is equivalent to 2.4 hours of sleep loss a week before or 0.2 hours of sleep loss every day for a week.
A more severe, but not uncommon scenario of losing an hour of sleep every day for a week is equivalent to losing 4.75 hours of sleep yesterday or 11.2 hours of sleep one week ago. 
On average, this loss in sleep debt was associated with a 9.5\% (6.2 points) reduction in game performance, and a 9.0\% (\$317) reduction in hires for salespeople.




\begin{figure}[t]
    \centering
    \subfigure[Game Grade]{
        \includegraphics[width=0.5\linewidth]{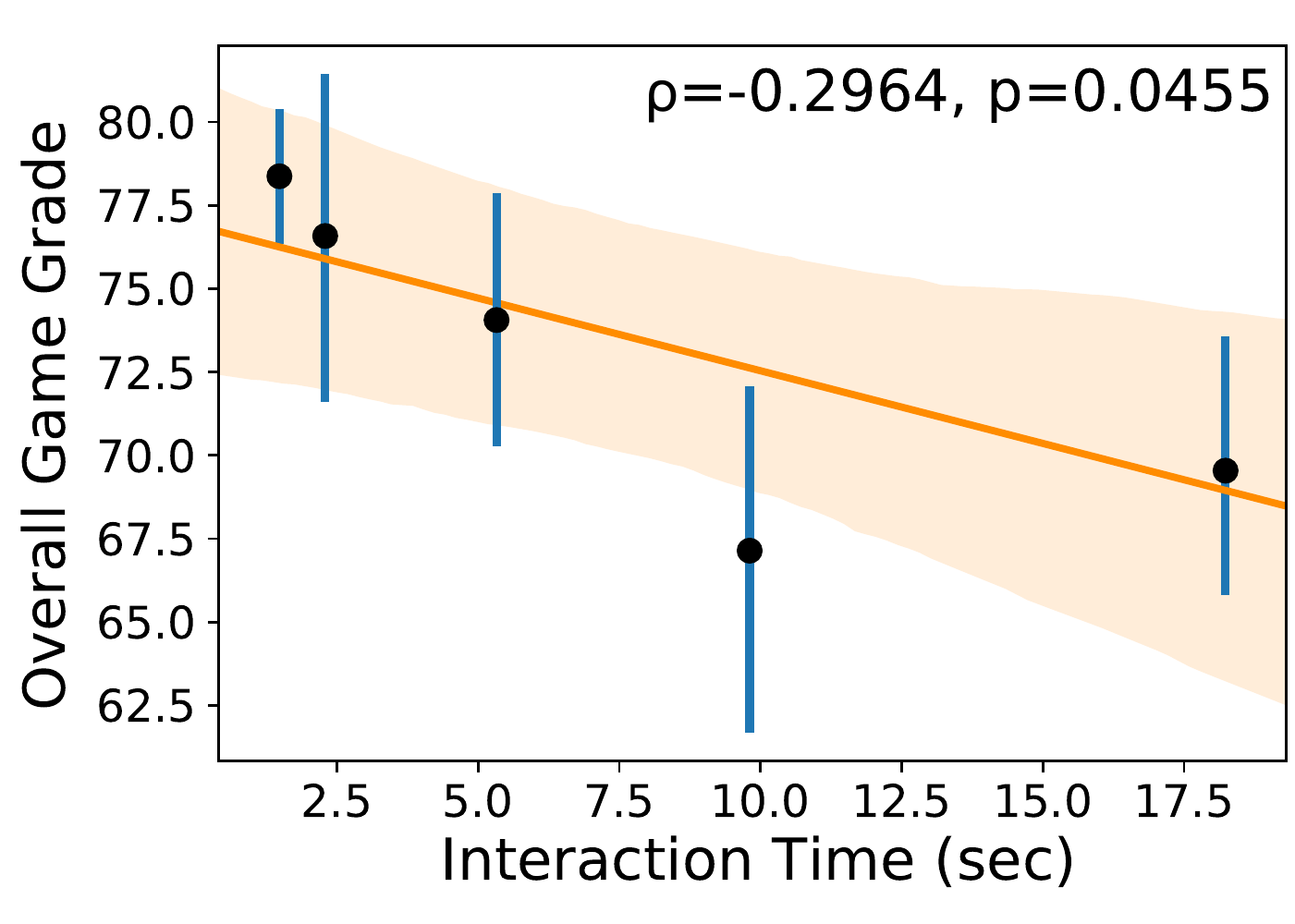}
        \label{fig:app-vs-job-perf-a}
    }%
    \subfigure[Hires Per Hour]{
        \includegraphics[width=0.5\linewidth]{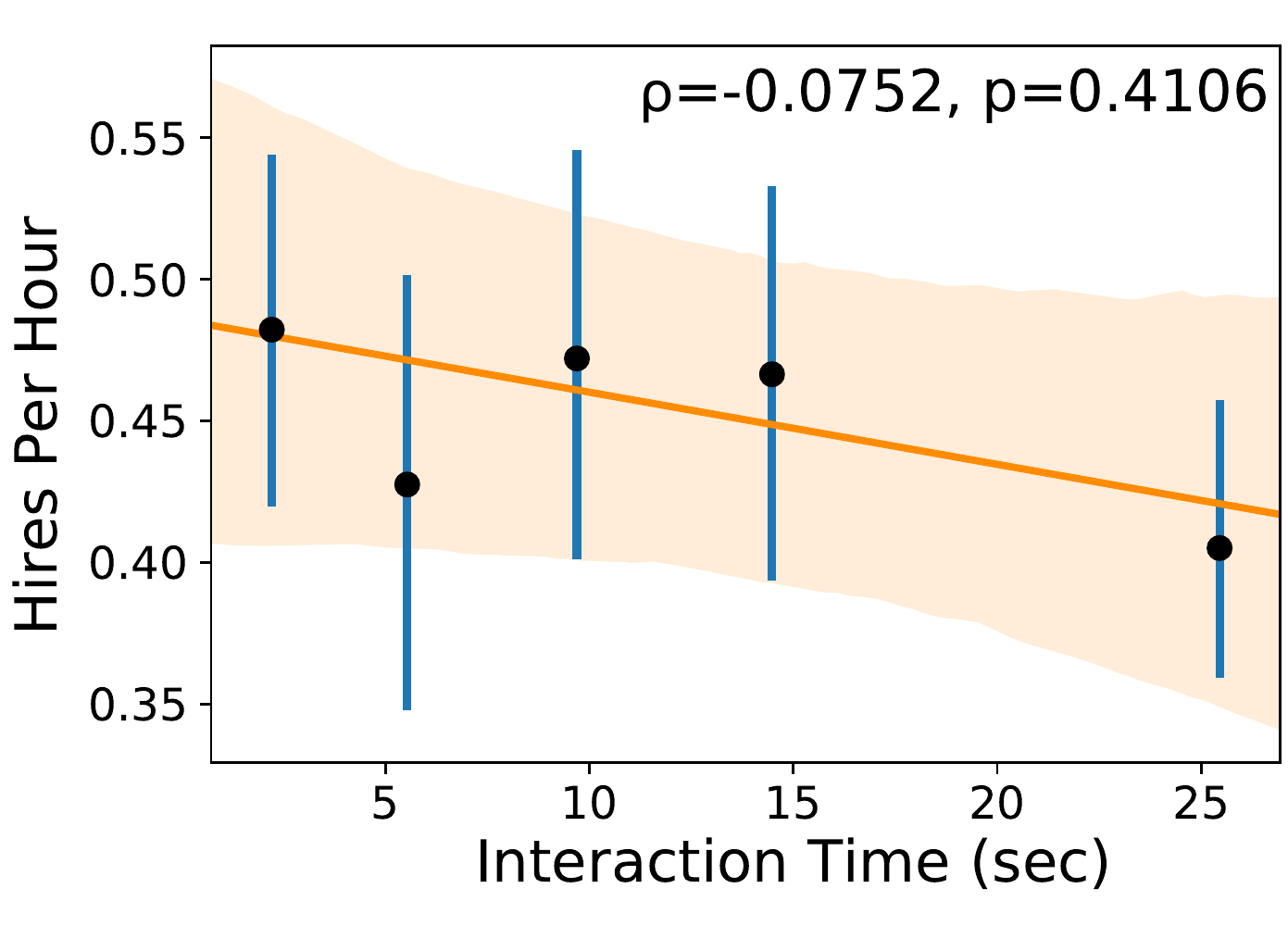}
        \label{fig:app-vs-job-perf-b}
    }
    \vspace{-3.00mm}
    \caption{Regression plots showing the effect sizes between app-based performance and (a) overall game grade and (b) hires per day. App interaction time is sensitive to the job performance of the athletes, but not the salespeople.}
    \label{fig:app-vs-job-perf}
    \vspace{-3.00mm}
\end{figure}

\subsection{RQ.2: The Relationship Between App Interaction Time and Job Performance}
\label{sec:rq2}
\final{Having supported the hypothesis that better sleep behavior is correlated with heightened job performance, we now explore the possibility of leveraging passively captured app interaction data as a non-invasive indicator of job performance.
We investigate this question on the basis that app-based performance provides an \textit{in-situ} measurement of psychomotor and cognitive function that may be easier to track than sleep behavior or job performance itself.}

\subsubsection{Analysis Procedure}
To examine whether app interaction time could serve as a non-invasive indicator of job performance, we calculate the correlation between these two data sources.
We also fit least squares models between app interaction time and job performance metrics to determine effect sizes.
The salespeople and athletes contributed data from 122 and 46 unique days with both app interaction and job performance measurements; note that this is a many-to-one relationship since participants frequently interacted with their app multiple times in the same day.

\begin{table*}[] 
    \centering
    \hspace*{-2.5cm}
    \begin{tabular}{c|*3{>{\centering\arraybackslash}m{0.75in}}|*3{>{\centering\arraybackslash}m{0.7in}}|}
    \cline{2-7}
    & \multicolumn{3}{c|}{ \textbf{Raw Sleep Data} } & \multicolumn{3}{c|}{ \parbox[b]{0.31\linewidth}{\centering \vspace{0.1cm} \textbf{Per Person \\ Z-Normalization of Sleep Data}}}\\
    \cline{2-7}
     & 
    Time-in- & 
    Sleep & 
    Sleep & 
    Time-in- & 
    Sleep &
    Sleep \\
    & 
    Bed & 
    History & 
    Debt & 
    Bed & 
    History &
    Debt \\
    \hline
    
    \multicolumn{1}{|c|}{\textbf{Interaction}} &
    
    
    
    
     \textbf{-0.015} &  \textbf{-0.055} &  \textbf{-0.095} & 0.006 & -0.012 & -0.010 \\

    \multicolumn{1}{|c|}{\textbf{Time}}& 
    

    
    
    \textbf{($p$=0.049)} & \textbf{($p$=}\bpvalue{3.9}{11}\textbf{)} &  \textbf{($p$=}\bpvalue{5.2}{30}\textbf{)} & ($p$=0.483) & ($p$=0.140) & ($p$=0.230) \\
    \hline

    \end{tabular}
    \hspace*{-2cm}
    \caption{Spearman correlation coefficients between sleep behavior and app-based performance. P-values are provided in parentheses; results with p-value < 0.05 are shown in bold.}
    \label{tab:spearman-corr-app-interaction}
    \vspace{-4.00mm}
\end{table*}
\newcolumntype{P}[1]{>{\centering\arraybackslash}p{#1}}

\begin{table}[t]
\centering
\begin{tabular}{|p{6cm}|P{1.75cm}|} \hline
\textbf{\studytwo Statistics} & \textbf{Participants} \\ \hline
Number of participants & 274 \\ \hline
Total nights of sleep tracked with app-based performance measurements & 7,195 \\ \hline
Total nights of sleep tracked & 30,618 \\ \hline
Total number of transitions between screens & 16,336 \\ \hline
Total number of times app was opened & 11,140 \\ \hline
Nights of sleep tracked per user (avg $\pm$ std) & 109.2 $\pm$ 91.81 \\ \hline
Time-in-bed in hours (avg $\pm$ std) & 7.338 $\pm$ 1.628 \\ \hline
Days of app use per user (avg $\pm$ std) & 43.68 $\pm$ 46.48 \\ \hline
\final{App interaction time in seconds (avg $\pm$ std)} & \final{10.14 $\pm$ 10.02} \\ \hline
\end{tabular}
\caption{Summary statistics for our dataset in \studytwo ~\final{after the filtering described in Section~\ref{sec:datasets-filtering}}.}
\label{tab:study2-dataset}
\vspace{-7.00mm}
\end{table}

\subsubsection{Results}
Figure \ref{fig:app-vs-job-perf} shows real-world job performance against app interaction time for those participants.
App interaction time was not found to be significantly correlated with the number of hires the salespeople made ($\rho$=-0.0752, $p$=0.411).
A significant correlation was found between app interaction time and the athletes' game grade ($\rho$=-0.296, $p$=0.0455). 
The effect size shows that athletes who were 10 seconds faster in their app interaction time had an average of 5 more points in game grades.
Our app interaction metric is partly related to reaction time, so the discrepancy between athletes and salespeople in this analysis may be because the athletes' day-to-day activities require rapid, precise reactions; the salespeople's activities, on the other hand, are typically more forgiving with respect to psychomotor function.
Another explanation could be that PFF includes contextual information, such as whether the opponent presented a favorable matchup during a game; the number of hires a salesperson can make in a given day is more dependent upon external factors (e.g., customer needs, health of the economy).



\section{\studytwo: General Population}
\label{sec:study2}

The previous study revealed statistically significant correlations between our app-based performance measurement and athletic job performance, but not salesperson job performance.
These findings highlight the fact that jobs can be extremely diverse (e.g., unique skill requirements and methods of rating performance), making it challenging to test the generalizability of our findings even further.
If app interaction time is truly an indicator of performance, it should be sensitive to factors that are known to impact psychomotor function.
Therefore, we conducted an exploration on a broader population of 274 participants to support the idea that app-based performance truly captures aspects of a person's psychomotor and cognitive function.
Using the PVT, sleep researchers have demonstrated that psychomotor and cognitive function improve with better sleep behavior~\cite{Rajdev2013,Ramakrishnan2012}.
Separately, computing researchers have shown that the timing between interaction events in a desktop or smartphone can be an indicator of psychomotor and cognitive function~\cite{Althoff2017,Vizer2009}.
Our third and final research question (\textbf{RQ.3}) aims to join these two bodies of literature.
Table~\ref{tab:study2-dataset} shows the summary statistics of the dataset used for this analysis after post-processing.

\subsection{RQ.3: The Relationship Between App Interaction Time and Sleep Behavior}
\label{sec:rq3}

\subsubsection{Analysis Procedure}
It is well established that psychomotor and cognitive function vary throughout the day due to circadian rhythms homeostatic sleep drive, and sleep inertia, collectively forming the three-process model of sleep~\cite{Akerstedt1997,Althoff2017,Matchock2009,Goel2013}.
Any performance indicator should therefore be sensitive to variations of time and sleep.
To examine whether this is the case for our app-based performance metric, we evaluate the relationship between app interaction time and four different measures: time of day, time since wake-up, sleep debt, and sleep history.
Beyond calculating the correlation between these two data sources, we also create a generalized additive model similar to the one proposed by Althoff et al.~\cite{Althoff2017} to characterize app interaction time as a function of sleep behavior and time of day.
We extend this model by incorporating random effects intercepts for each user, which not only accommodates user-specific performance baselines, but also accounts for device-specific effects like the rendering capabilities of the user's smartphone.
If some participants took certain medications, regularly napped, or consistently consumed high quantities of caffeine, these confounds would be adjusted through the random intercepts as well.
Our participants logged 7,195 nights of sleep that were paired with at least one app interaction event during the same day.

\begin{figure}[t]
    \centering
    \subfigure[Time-in-bed]{
        \includegraphics[width=0.5\linewidth]{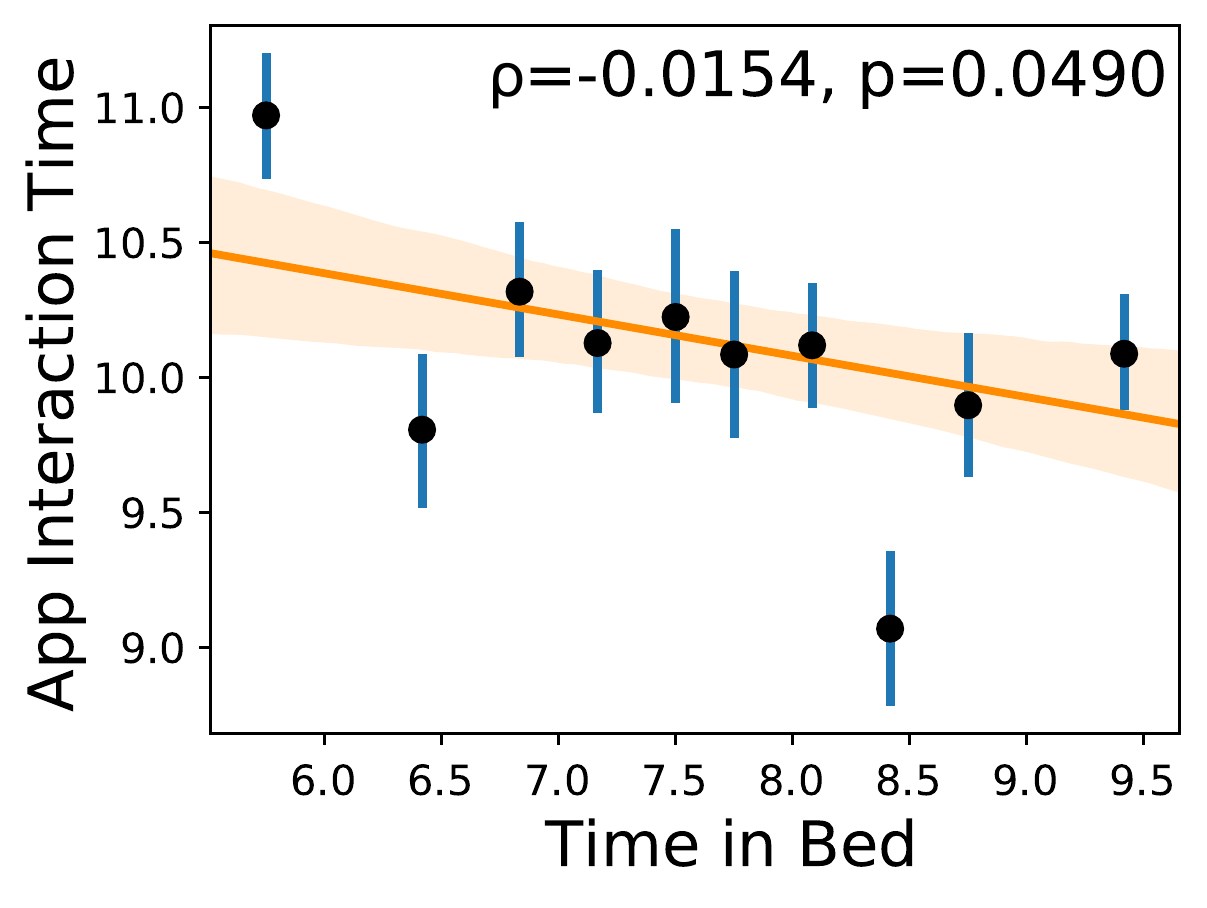}
        \label{fig:sleep-app-stat-sig-a}
    }
    \subfigure[Sleep History]{
        \includegraphics[width=0.5\linewidth]{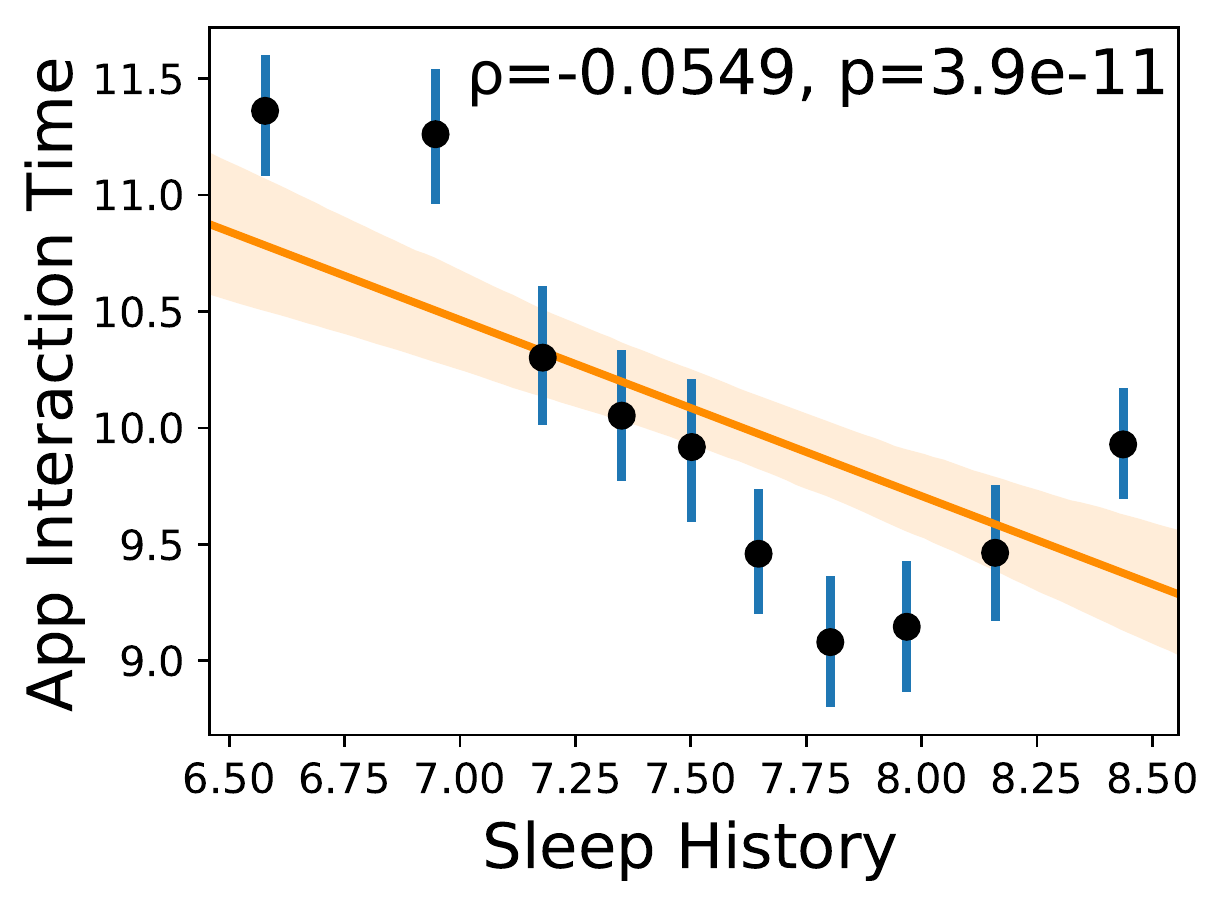}
        \label{fig:sleep-app-stat-sig-b}
    }%
    \subfigure[Sleep Debt]{
        \includegraphics[width=0.5\linewidth]{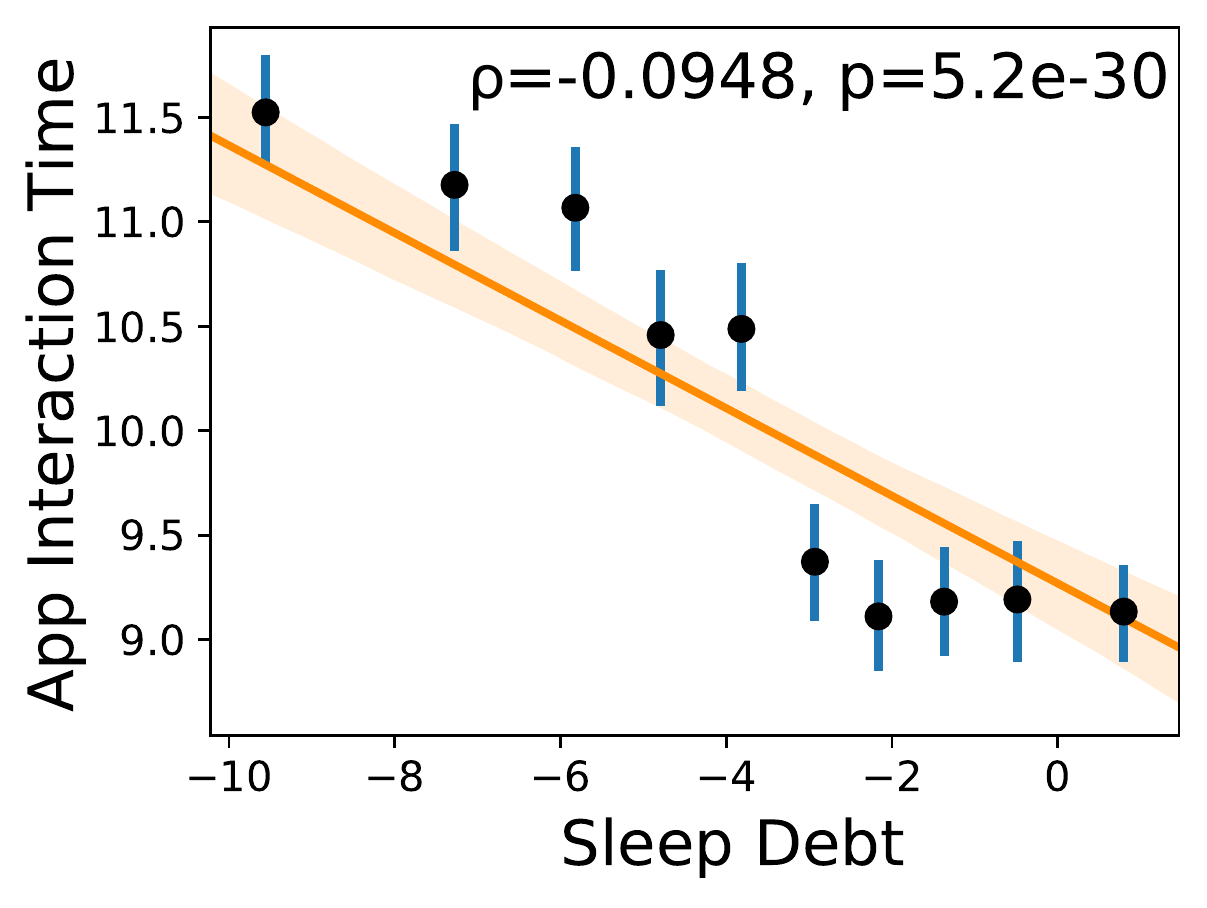}
        \label{fig:sleep-app-stat-sig-c}
    }
    \vspace{-3.00mm}
    \caption{Regression plots showing the effect sizes for the statistically significant results from Table \ref{tab:spearman-corr-app-interaction}. App interaction time is sensitive to many sleep metrics: (a) time-in-bed, (b) sleep history, and (c) sleep debt.
    }
    \label{fig:sleep-app-stat-sig}
    \vspace{-5.00mm}
\end{figure}

\subsubsection{Results}
Table~\ref{tab:spearman-corr-app-interaction} summarizes the correlation coefficients between sleep behavior and app interaction time. 
Time-in-bed ($\rho=-0.0154$ $p=0.049$), sleep history ($\rho=-0.0549$, $p=3.9\times10^{-11}$), and sleep debt ($\rho=-0.0948$, $p=5.2\times10^{-30}$) had negative correlations with app interaction time; in other words, participants with better sleep behaviors had faster app interaction times.
Although the correlation coefficients on individual performance are rather small due to significant variation within and across participants, these estimates align with findings from previous work \cite{Althoff2017, Lo2016, VanDongen2003}.
When averaging the app interaction times of samples within a certain sleep metric bin, the effect sizes are practically meaningful and span differences of up to 2.5 seconds.
For example, as shown in Figure~\ref{fig:sleep-app-stat-sig-c}, one hour less of sleep debt was associated with app interactions times that were 0.175 seconds slower than the average. 
Another way to frame this effect size is that one hour less of daily sleep over the past week was associated with app interaction times that were 0.72 seconds slower.
As before, the cumulative sleep behavior metrics exhibited stronger correlations than total time-in-bed; however, app-based performance correlated better with non-normalized sleep behavior metrics. 
This result suggests that the minimal complexity of the app interaction task engendered less variance across individuals. 
Moreover, we found that extended sleep does not improve psychomotor performance. 
Figure~\ref{fig:sleep-app-stat-sig-b} shows that app interaction time was fastest when individuals had an average of 7.75 hours of daily sleep over the past week.
Similar U-shaped relationships have also been reported in previous work on psychomotor performance~\cite{Althoff2017} and other outcomes (e.g., mortality~\cite{kripke1979short}).

\begin{figure}[t]
    \centering
    \subfigure[Accounting for Sleep History]{
    \hspace{-4mm}
    \label{fig:app-perf-a}
    \includegraphics[width=0.51\linewidth]{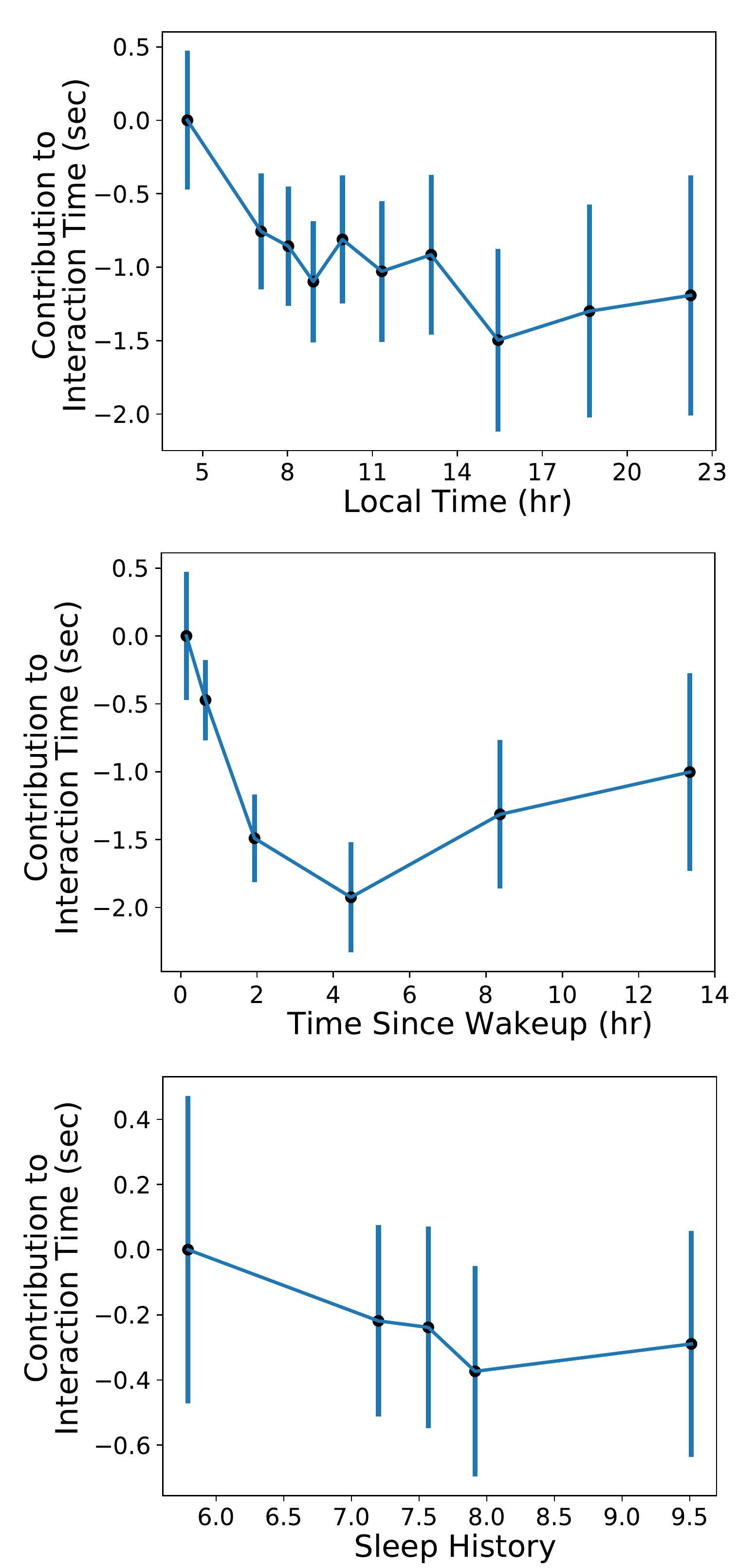}}%
    \subfigure[Accounting for Sleep Debt]{
    \label{fig:app-perf-b}
    \includegraphics[width=0.51\linewidth]{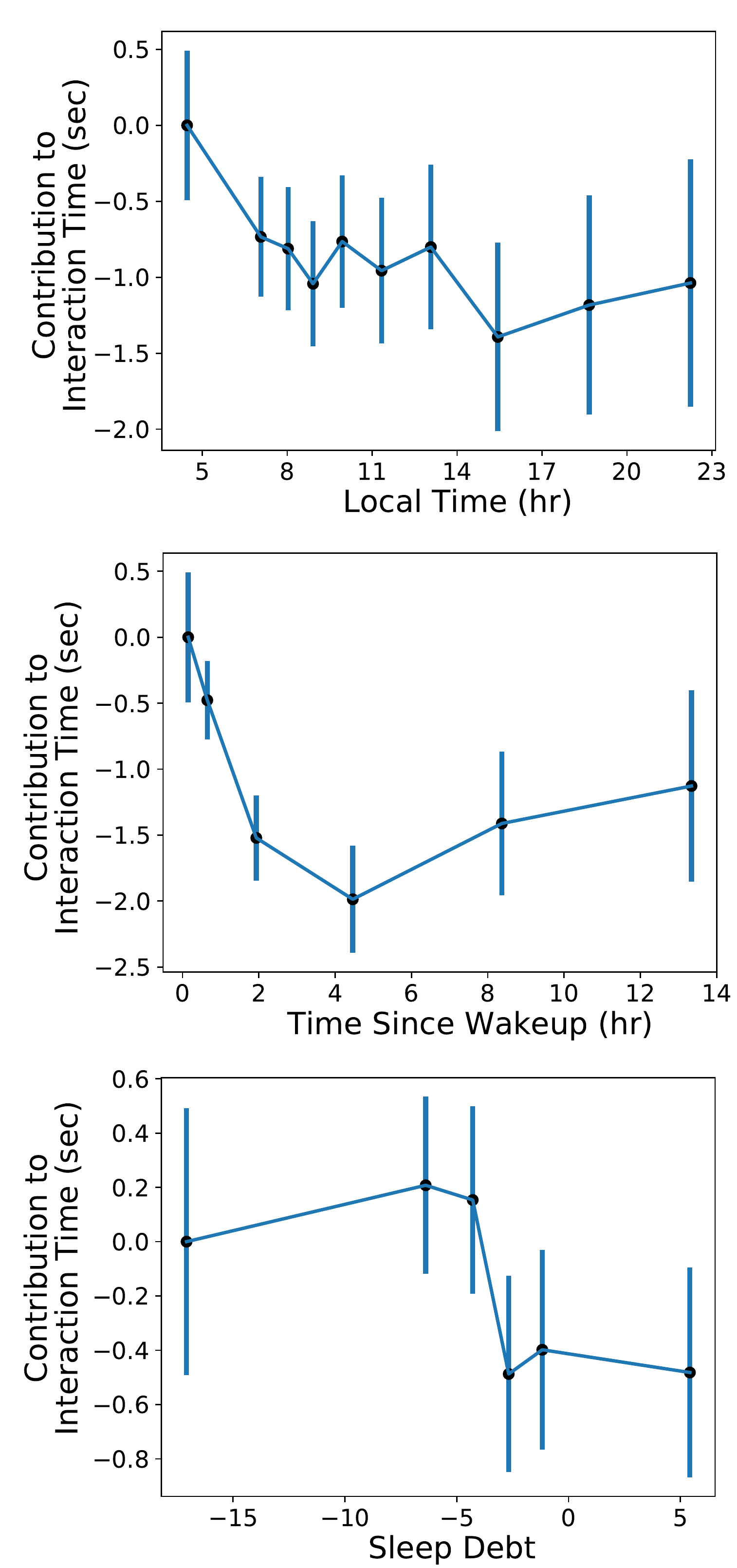}}
    \vspace{-3.00mm}
    \caption{Generalized additive models of app interaction time accounting for (a) sleep history and (b) sleep debt. In both cases, the models account for (top row) the local time in the participant's time zone, (center row) time since wake-up, and (bottom row) sleep behavior. These models show that app interaction time is sensitive to sleep behaviors including circadian rhythm, time awake, and cumulative sleep metrics. Both models include random intercepts for each participant, and standard errors are shown.
    }
    \label{fig:app-perf}
    \vspace{-5.00mm}
\end{figure}

Since we found statistically significant correlations between cumulative sleep behavior metrics and job performance, we used sleep history and sleep debt in generalized additive models.
Figure \ref{fig:app-perf} shows the variation of app interaction time as a function of time of day, time since wake-up, and the aforementioned metrics. 
We find that app interaction times are slowest at night and fastest between 3-6 PM; the difference between those extremes is approximately 1.5 seconds.
\final{Note that the relationship between app interaction time and time of day (Figure~\ref{fig:app-perf}, top) generally aligns with circadian rhythm processes as measured through controlled sleep studies~\cite{Dijk1992,Borbely2016,Akerstedt1997}.}
Our results also align with the chronobiological process of sleep inertia~\cite{Akerstedt1997} since participants had slower app interaction times within one hour of waking up (Figure~\ref{fig:app-perf}, middle).
App interaction time decreases in the first six hours after wake-up and then begins to increase again, consistent with both the chronobiological process of homeostatic sleep drive~\cite{Borbely2016} and previous work examining click speeds in search engines~\cite{Althoff2017}.
App interaction time increased by an average of 0.4 seconds when sleep history improved from 6 to 8 hours, and app interaction time increased by 0.5 seconds beyond the threshold of -5 sleep debt hours. 
In other words, when participants lost one hour of sleep daily for a week, they exhibited app interaction times that were \final{5\% (0.5 seconds)} slower.
Note that this estimate is slightly less than the estimate of 0.72 seconds in Figure~\ref{fig:sleep-app-stat-sig-c}. 
The difference between the two estimates is explained by the fact that the generalized additive model controls for the impacts of circadian rhythm, homeostatic sleep drive, and sleep inertia, as well as participant-specific baselines through random effects.
\section{Discussion}
Establishing the relationship between sleep behavior and job performance has been a challenge in the past due to the difficulty in collecting objective measures in real-world settings.
By taking advantage of ubiquitous sleep-tracking technology and the increasing desire within companies to evaluate job performance through data, our research signifies a major step towards understanding this relationship.
We demonstrate that an app-based performance metric is correlated with both job performance metrics and sleep behaviors in a way that is consistent with sleep biology.
This highlights an interesting opportunity for future assessments of sleep and performance in uncontrolled settings.
Below, we describe the implications and limitations of our work.

\subsection{Opportunities for Passive Sensing}
\label{sec:discussion-passive}
The PVT has been used to measure psychomotor and cognitive function in the wild~\cite{Abdullah2016}; however, the PVT can be disruptive if deployed at inopportune moments.
Other prior work has required participants to adhere to a strict sleep schedule in order to measure the effects of sleep on behavior~\cite{Lo2016,VanDongen2003}.
In our work, we found that our instantiation of app-based performance was correlated with both better sleep behavior and athletic job performance, suggesting the potential power of a passive, nonintrusive performance indicator.
Passive sensing through ubiquitous technologies like smartphones can enables continuous data collection for the study of populations that have traditionally been difficult to recruit to controlled studies. 

We restricted our correlation analysis of app-based performance to comparable interactions within the sleep-tracking app that started from the home screen and involved single touches; however, not all interactions are created equal, nor does app interaction time tell the whole story about how the user is engaging with the app's content.
Some screens require more time to process than others, and longer processing times may indicate that the user is engaging more with the displayed information.
Understanding how app interaction time is a function of on-screen content could be explored further to enable more robust measurements.
Beyond app interaction time, comparable performance metrics have also been elicited through other interactions like typing and web browsing~\cite{Thirkettle2018,Althoff2017,Vizer2009}.
Responses to alarms and notifications could also provide more natural opportunities for capturing app-based performance in the future.

\subsection{Implications for Sleep-Tracking Apps}
\label{sec:discussion-recommendations}
One design recommendation that we propose for sleep-tracking apps involves personalized views of sleep metrics.
Many researchers have noted that sleep behaviors are unique according to genetic predisposition and chronotyping~\cite{Roenneberg2003,Allebrandt2010}.
Throughout our analyses, there were cases when normalizing sleep behavior metrics according to each user's history produced statistically significant correlations, but the same was not true for the raw data.
Presenting raw values in combination with data that is scaled relative to the individual could provide useful insights to users in the future.
Because sleep quality is subjective and not well-defined \cite{Harvey2008,Ravichandran2017}, future apps could also allow users to explore what sleep metrics matter to their perceived sleep quality.
In fact, we posit that job performance may be influenced by a person's perception of their own sleep quality, so our research may inform ways of exploring this matter in the future.

Finally, lapses in sleep tracking and the resulting lack of data are an important consequence of real-world data collection that should be addressed.
Our dataset exhibited an extreme case of this issue since athletes can be away from home for at least 3-4 days at a time; nevertheless, travel is a regular occurrence for many people.
The cumulative sleep metrics in our dataset---sleep history and sleep debt---were most informative in our analyses related to sleep behavior.
We used the average time-in-bed of nearby nights for imputation when a participant skipped a night of sleep tracking (Section~\ref{sec:datasets-filtering}).
\final{Future work could explore other alternatives to imputation, such as improving generative models through deep learning~\cite{fang2020time,yoon2018gain} or multi-device sensing to remedy data gaps~\cite{Ko2015}.}

\subsection{Additional Context Information}
\label{sec:discussion-context}
PFF game grades are able to incorporate context because they are assigned by experts who watch the games and understand the athletes' match. 
Our other data streams, however, lacked such context.
For example, the performance of salespeople depends on the demand of their goods and services.
Job performance in general is also a function of experience and division of labor.
Such information from managers and worker profiles could be incorporated for refined analyses in future work.

Sleep is known to be affected by a wide variety of factors: age~\cite{Yoon2003,Dijk1999,Feinberg1974}, ambient light~\cite{Kozaki2005}, caffeine intake~\cite{Landolt1995}, and diet~\cite{Halson2008}, to name a few.
The effect of travel between time zones (2--3 hour difference) has not been shown to significantly impact sleep~\cite{Richmond2007}, but an effect has been demonstrated on athletic performance~\cite{Jehue1993}.
Measuring these factors through sensors and accounting for their effects in statistical analyses could improve evidence of links between sleep behavior, job performance, and app usage.

\subsection{Limitations}
\label{sec:discussion-limitations}
Our dataset included participants from a bankruptcy law firm consultancy and the NFL, which allowed us to compare two populations with distinct job demands whose job performance can be quantified effectively.
In both cases, we were able to identify sleep behavior metrics that correlated with job performance; however, the correlations manifested in different sleep behavior metrics (e.g., sleep debt for salespeople, personalized sleep history for athletes) (Section~\ref{sec:rq1}.
Beyond the discrepancy between the two groups' job demands, the differences in results can also be attributed to idiosyncrasies within the job performance metrics themselves.
For the salespeople, the number of hires an employee is able to make may depend on the state of the economy and the rate of bankruptcy in the country.
For the athletes, the subjective nature of the expert's grades can manifest in anchoring effects towards common values~\cite{Tversky1974}.
We use rank-based correlation methods and per-person normalization to account for some of these idiosyncrasies (Section~\ref{sec:rq1}), but future work should explore and compare alternative sources of job performance data.
Furthermore, an exciting avenue of research may entail the creation of a job performance metric that generalizes across different careers.

Although salespeople and athletes have very different job demands, they do not cover the entire spectrum of careers.
Each profession has its own demands and may not overlap with either of the ones that were included in our study.
There was also an element of selection bias in our participant pool; the people who enrolled in our observational study may have been more excited to track their sleep and interact with the app than the average person, producing inflated app engagement measurements.
Similarly, the observational and correlational nature of our data preclude us from making causal inferences.
Learning about how our findings may generalize to other populations remains an area of future work.

Lastly, there are many confounds that could have affected our datasets.
People have unique habits that affect their sleep behavior and job performance~\cite{Kozaki2005,Landolt1995,Halson2008}. 
Unique smartphone parameters like clock speed or operating system throttling due to current battery level affect app interaction time.
We addressed within-person confounds as much as possible via statistical methods.
For our correlational analyses, we examined both raw and per-person normalized sleep behavior metrics (Section~\ref{sec:rq1}, \ref{sec:rq2}, \ref{sec:rq3}).
For our generalized additive model of app interaction time against sleep behavior and time of day, we utilized random effects intercepts to accommodate for performance baselines, habits, and device specifications specific to each participant (Section~\ref{sec:rq3}). 
These steps helped us account for confounds that existed throughout a participant's enrollment in the study, including regular medication intake, naps, or caffeine consumption.
\section{Conclusion}
Many people recognize that improving sleep behavior benefits job performance, but the precise relationship between the two has been difficult to capture and quantify in the past.
Our study advances the literature in this space by providing a correlational analysis between objectively measured sleep behavior metrics from a mattress sensor and job performance metrics from a bankruptcy law firm and the NFL.
Our findings suggest that establishing good sleep behaviors over extended periods is more important to job performance than simply getting a good night's sleep one day prior. 
We also found evidence that passively captured app interaction metrics can serve as a useful indicator for some job performance and sleep measures, thereby highlighting another mechanism through which researchers can collect relevant psychomotor and cognitive performance measures at scale. 
It is our hope that our work inspires researchers to examine \textit{in-situ} sleep behaviors and performance measures across diverse contexts to further develop our understanding of human performance.

\begin{acks}
This research has been supported in part by NSF grant IIS-1901386, Bill \& Melinda Gates Foundation (INV-004841), the Allen Institute for Artificial Intelligence, and a Microsoft AI for Accessibility grant.
\end{acks}

\bibliographystyle{ACM-Reference-Format}
\balance
\bibliography{chapters/references}


\end{document}